\documentclass[a4paper,11pt]{article}

\usepackage[english]{babel}
\usepackage{geometry}
\usepackage{amsmath, amssymb, amsfonts} 
\usepackage{mathrsfs} 
\usepackage{graphicx}
\usepackage{tikz}
\usepackage{titlesec}
\usepackage{fancyhdr}
\usepackage{xcolor} 
\definecolor{linkblue}{RGB}{0, 0, 255} 
\usepackage[colorlinks=true, linkcolor=linkblue, citecolor=linkblue, urlcolor=linkblue]{hyperref}
\usepackage{amsthm} 
\usepackage{cleveref}
\usepackage{makecell} 

\numberwithin{definition}{section} 

\definecolor{commentblue}{RGB}{0,90,200}
\usepackage[textsize=small, disable]{todonotes}

\usetikzlibrary{automata,positioning}
\usepackage{enumitem}
\setlist[itemize]{label=--}

\usepackage{lscape} 
\usepackage{datetime} 

\usepackage[font=small, labelfont=bf]{caption}

\usepackage{booktabs}    
\usepackage[table]{xcolor}
\usepackage{tabularx}    

\usepackage[utf8]{inputenc}
\usepackage[T1]{fontenc}
\usepackage{orcidlink}
\newcommand{\orcid}[1]{\orcidlink{#1}}
\usepackage{microtype}       
\usepackage[percent]{overpic}

\usepackage{multirow}
\usepackage{array}

\usepackage{pgfplots}
\pgfplotsset{compat=1.17}

\usepackage[style=numeric-comp, sorting=nyt, maxcitenames=2, date=year, doi=false, isbn=false, url=false, eprint=false, giveninits=true]{biblatex}
\usepackage{subcaption}
\usepackage{float}
\DeclareNameAlias{author}{last-first}

\DeclareFieldFormat[article]{title}{#1}                  
\DeclareFieldFormat[article]{journaltitle}{\textit{#1}}  
\DeclareFieldFormat[article]{volume}{\textbf{#1}}        
\DeclareFieldFormat[article]{number}{(#1)}               
\DeclareFieldFormat[article]{pages}{#1}                  

\renewbibmacro{in:}{}

\crefname{figure}{Fig.}{Figs.}  
\crefname{table}{Tab.}{Tabs.}   
\crefname{equation}{Eq.}{Eqs.}  
\crefname{definition}{Definition}{Definitions} 
\crefname{theorem}{Theorem}{Theorems} 

\usepackage[version=4]{mhchem}
\usepackage{stmaryrd}

\usepackage[export]{adjustbox}
\usepackage{multirow}

\begin{document}

\title{Sensor Informativeness, Identifiability, and Uncertainty in Bayesian Inverse Problems for Structural Health Monitoring}

\author{
    Tammam Bakeer\thanks{Independent Researcher, Dresden, Germany. Corresponding author: \href{mailto:mail@bakeer.de}{mail@bakeer.de}} 
    \orcid{0000-0001-6475-3907}\,
    \quad
    Max Herbers\thanks{Technische Universität Dresden, Institute of Concrete Structures, Germany}\,\,%
    \orcid{0000-0002-2187-1652}
    \quad
    Steffen Marx\footnotemark[2]\,\,%
    \orcid{0000-0001-8735-1345}
}

\date{\today}

\maketitle

\begin{abstract}
In Structural Health Monitoring (SHM), the recovery of distributed mechanical parameters from sparse data is often ill-posed, raising critical questions about identifiability and the reliability of inferred states. While deterministic regularization methods such as \emph{Tikhonov} stabilise the inversion, they provide little insight into the spatial limits of resolution or the inherent uncertainty of the solution. This paper presents a Bayesian inverse framework that rigorously quantifies these limits, using the identification of distributed flexural rigidity from rotation (tilt) influence lines as a primary case study. \emph{Fisher information} is employed as a diagnostic metric to quantify sensor informativeness, revealing how specific sensor layouts and load paths constrain the recoverable spatial features of the parameter field.

The methodology is applied to the full-scale openLAB research bridge (TU Dresden) using data from controlled vehicle passages. Beyond estimating the flexural rigidity profile, the Bayesian formulation produces credible intervals that expose regions of practical non-identifiability, which deterministic methods may obscure. The results demonstrate that while the measurement data carry high information content for the target parameters, their utility is spatially heterogeneous and strictly bounded by the experiment design. The proposed framework unifies identification with uncertainty quantification, providing a rigorous basis for optimising sensor placement and interpreting the credibility of SHM diagnostics.

\end{abstract}

\vspace{1cm} 

\newpage

\section{Introduction}
\label{sec:intro}

The sudden collapse of the Carola Bridge in Dresden in September 2024, attributed to stress–corrosion cracking and tendon degradation, highlighted the vulnerability of aging prestressed concrete bridges and the urgent need for reliable diagnostic methods \autocite{marxEinsturzCarolabrueckeDresden2025,schachtEinsturzCarolabrueckeDresden2025}. Beyond reaffirming the importance of structural health monitoring (SHM), this event underscored the necessity of quantifying the \emph{credibility} of monitoring-based inferences so that decision makers can judge the reliability of estimated structural capacity.

Bridges worldwide are increasingly operating beyond their intended service lives and under growing traffic demands, which heightens the need for reliable indicators of structural capacity. Because direct measurement of internal structural condition is rarely feasible, latent states must instead be inferred from indirect and typically noisy observations. This task constitutes an \emph{inverse problem}, where the objective is to recover unobservable structural properties from measurable response data. Within this framework, \emph{flexural rigidity} is of particular importance, as its progressive degradation often precedes observable serviceability or safety issues. Accordingly, the distributed flexural rigidity profile has become a primary target for SHM, supplying essential input for load-rating procedures and the calibration of digital twin models \autocite{hesterIdentifyingDamageBridge2020,huseynovBridgeDamageDetection2020}.

Flexural rigidity aggregates material stiffness and section geometry into the quantity that dictates how loads map to deflection and rotation, so departures from its nominal profile immediately reflect fabrication tolerances, damage, or material degradation \autocite{lucchinettiMeasuringFlexuralRigidity2002}. For bridges ageing under fatigue, corrosion, and overloads, spatially resolved estimates of flexural rigidity underpin damage localisation, load rating, and maintenance planning because they expose flexural rigidity losses before serviceability or strength limits are reached \autocite{youDistributedBendingStiffness2023}. Consequently, accurately inferring flexural rigidity from noisy field data is not an auxiliary task but the primary enabler of actionable diagnostics for bridge owners, even though the inverse problem remains highly ill-posed \autocite{lucchinettiMeasuringFlexuralRigidity2002}. In prestressed concrete bridges, tendon corrosion, prestress loss, bond degradation, and stress–corrosion cracking have all been documented to diminish flexural rigidity, causing detectable changes in deformation patterns well before ultimate capacity is compromised \autocite{
podolnyCorrosionPrestressingSteels1992,
fuzierDurabilityPosttensioningTendons2006,
luNewEmpiricalModel2023,
wangStrandCorrosionPrestressed2023}. Spatially resolved estimates of effective flexural rigidity therefore provide an early and actionable diagnostic for asset managers.

Rotation and tilt measurements have become practical sensing modalities because modern inclinometers and inertial sensors resolve microradian changes without external references \autocite{mcgeownUsingMeasuredRotation2021,zhouNovelElastomerBasedInclinometer2022}. Rotations are largest near supports, easing installation, and comparative laboratory and field studies show that \emph{rotation influence lines} remain sensitive to local reductions in flexural rigidity across diverse load configurations \autocite{hesterIdentifyingDamageBridge2020,huseynovBridgeDamageDetection2020,faulknerTrackingBridgeTilt2020,wangNovelBridgeDamage2023}.

Classical inverse strategies reconstruct stiffness profiles from deflection influence lines using discretisation, smoothing, and adaptive unit-load schemes \autocite{youDistributedBendingStiffness2023}. Earlier formulations also inferred flexural rigidity from sparse static deflections, but only under strong regularisation to temper the \emph{ill-posed Fredholm equations} \autocite{lucchinettiMeasuringFlexuralRigidity2002}. Even when effective, deflection-based schemes often require dense instrumentation, a stable reference frame, and aggressive filtering, which limits practicality for in-service bridges \autocite{lucchinettiMeasuringFlexuralRigidity2002,youDistributedBendingStiffness2023}.

Despite these advances, most existing contributions either treat rotation as a qualitative damage indicator or reconstruct flexural rigidity within deterministic inverse frameworks, most notably formulations that estimate distributed flexural rigidity from \emph{rotation-derived} deformation influence lines using iterative (multi-parameter) \emph{Tikhonov regularisation} \autocite{zeinaliFrameworkFlexuralRigidity2017,zeinaliImpairmentLocalizationQuantification2018}. However, fundamental questions remain regarding the identifiability of flexural rigidity, the amplification of measurement noise by ill-posed operators, and the principled quantification of uncertainty in the recovered fields \autocite{hesterIdentifyingDamageBridge2020,huseynovBridgeDamageDetection2020}. Such quantification is essential if SHM outcomes are to be trusted for safety-critical decisions and incorporated into value-of-information analyses for bridge management \autocite{thonsValueMonitoringInformation2018,kamariotisValueInformationVibrationbased2022}.

We address these gaps by developing a \emph{Bayesian inverse framework} for identifying distributed flexural rigidity from rotation measurements. The formulation incorporates engineering priors and a noise characterisation to obtain posterior distributions that make identifiability limits explicit. The Bayesian viewpoint interprets classical regularisation as the specification of a prior, clarifies the correspondence between estimator structure and data weighting, and yields credible intervals in addition to point estimates \autocite{stuartInverseProblemsBayesian2010,kaipioStatisticalComputationalInverse2006}.

Section~\S\ref{sec:1} formalises the structural response model and states the associated inverse problem for distributed $EI(x)$. We then construct the Bayesian inference workflow, assess sensor informativeness using Fisher information (\S\ref{subsec:fisher}), examine posterior uncertainty and identifiability (\S\ref{sec:uncertainty}), and demonstrate the methodology on the openLAB research bridge (\S\ref{sec:openlab}), concluding with implementation guidance for SHM practice. The resulting posterior distributions and credibility intervals delineate the limits of what rotation measurements can reveal, enabling responsible deployment of rotation-based diagnostics on ageing prestressed bridges.

\section{Forward Problem and Inverse Formulation}\label{sec:1}

Consider an \emph{Euler--Bernoulli beam} on $[0,L]$, simply supported at $x=0,L$, in the small-deflection regime. The (unknown) \emph{flexural rigidity} and \emph{compliance} are
\begin{equation}
EI(x)>0,\qquad v(x):=\frac{1}{EI(x)} .
\label{eq:v_def}
\end{equation}
Let $w(x)$ denote the transverse deflection; $\theta(x)=w'(x)$ and $\kappa(x)=w''(x)$ in the small-deflection \emph{Euler--Bernoulli theory}.
For a point load $P$ at $z\in[0,L]$,
\begin{equation}
\kappa(x)=\theta'(x)=\frac{M(x;z)}{EI(x)} .
\label{eq:curvature}
\end{equation}

Rotations (small-angle \emph{tilts}/\emph{inclinations}) are measured at stations $\mathcal S=\{r_1,\dots,r_R\}\subset(0,L)$ and for load positions $Z=\{z_1,\dots,z_K\}$. The measurement model is
\begin{equation}
y_{rk}=\theta(r;z_k)+\varepsilon_{rk}, \qquad r\in\mathcal{S},\; k=1,\dots,K .
\label{eq:meas}
\end{equation}

Here $\varepsilon_{rk}$ denotes zero-mean noise; the Gaussian noise model is specified in Section~\S\ref{sec:bayes}.

By \emph{Maxwell--Betti/Castigliano}, the rotation at $r$ under a load at $z$ is
\begin{equation}
\theta(r;z)=\int_0^L m_r(s)\,M(s;z)\,v(s)\,ds .
\label{eq:forward}
\end{equation}

Here, \(m_r(s)\) denotes the bending-moment influence function 
(the bending moment at coordinate \(s\) resulting from a unit couple applied at position \(r\)), 
and \(M(s;z)\) is the bending-moment distribution along the span 
generated by a point load of magnitude $P$ acting at \(x=z\).
The integration variable \(s\in[0,L]\) represents the spatial coordinate along the beam.

For a simply supported span, the bending moment influence function and the load-induced bending moment take the forms  
\begin{align}
m_r(s) \;&=\; -\tfrac{s}{L} \;+\; H(s-r), \\[6pt]
M(s;z) \;&=\; P\!\left[\tfrac{L-z}{L}\, s \, H(z-s) 
+ z\!\left(1 - \tfrac{s}{L}\right) H(s-z)\right],
\label{eq:mrM}
\end{align}
where \(H(\cdot)\) denotes the \emph{Heaviside function}, acting on the integration variable \(s\), used to represent the piecewise character of the relations in a compact form.

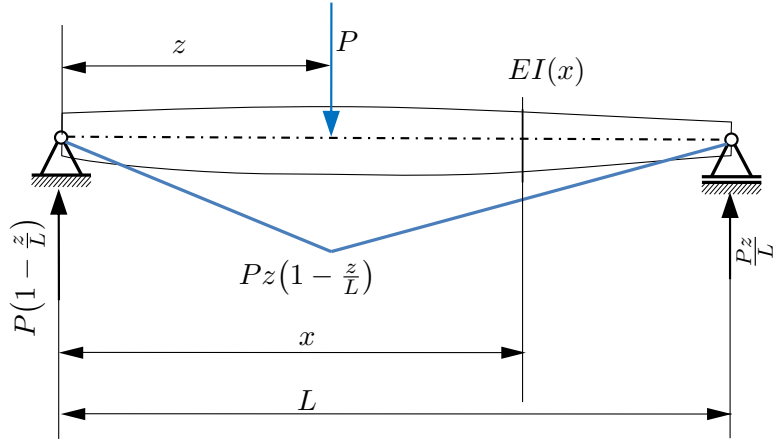
\begin{figure}[htbp]
\centering
\begin{tikzpicture}
  \node[anchor=south west, inner sep=0] (img) at (0,0)
    {\includegraphics[width=0.7\linewidth]{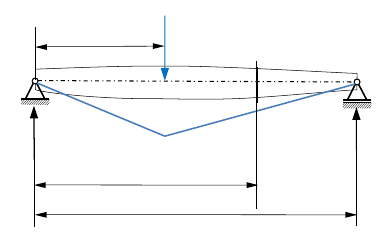}};

  \begin{scope}[x={(img.south east)}, y={(img.north west)}]
    \node at (0.24,0.850) {$z$};
    \node at (0.45,0.86) {$P$};
    \node at (0.7,0.8) {$EI(x)$};

    \node[rotate=90] at (0.05,0.4) {$P\!\left(1-\tfrac{z}{L}\right)$};
    \node[rotate=90] at (0.96,0.45) {$\tfrac{P z}{L}$};

    \node at (0.40,0.4) {$P z \!\left(1-\tfrac{z}{L}\right)$};

    \node at (0.4,0.28) {$x$};
    \node at (0.4,0.16) {$L$};

  \end{scope}
\end{tikzpicture}
\caption{Simply supported Euler--Bernoulli beam with span $L$, point load $P$ at $z$, bending moment diagram, and variable flexural rigidity $EI(x)=E(x)I(x)$.}
  \label{fig:beam_setup}
\end{figure}

Defining the \emph{kernel} (with respect to the unknown compliance) by
\begin{equation}
K(r,z;s):=m_r(s)\,M(s;z),
\end{equation}
the forward map is the \textit{first-kind Fredholm equation} \autocite{hansenDiscreteInverseProblems2010}
\begin{equation}
\theta(r;z)=\int_0^L K(r,z;s)\,v(s)\,ds,
\label{eq:kernel_form}
\end{equation}
which is linear (and compact) in $v(\cdot)$.

\subsection*{Discretization}
Partition $[0,L]$ into $N$ elements with endpoints $\{s_j\}_{j=0}^N$, lengths $\Delta s_j:=s_j-s_{j-1}$, and midpoints $\{\tilde{s}_j\}_{j=1}^N$. Define the element value of the (unknown) compliance as the \emph{element average}
\begin{equation}
\label{eq:v_avg}
 v_j := \frac{1}{\Delta s_j}\int_{s_{j-1}}^{s_j} v(s)\,ds,\qquad j=1,\dots,N.
\end{equation}
Using the partition to decompose the forward integral \eqref{eq:kernel_form} gives
\begin{equation}
\label{eq:element_sum}
\theta(r;z)
 = \sum_{j=1}^{N} \int_{s_{j-1}}^{s_j} K(r,z;s)\,v(s)\,ds
\;\approx\; \sum_{j=1}^{N} v_j\,\underbrace{\int_{s_{j-1}}^{s_j} K(r,z;s)\,ds}_{=:\,A_{(r,z),j}}.
\end{equation}
Evaluating $A_{(r,z),j}$ at the discrete load positions $z=z_k$ gives the matrix entries $A_{(r,k),j}$ used below. The approximation becomes exact when $v$ is piecewise constant on the elements and is first-order consistent under refinement.

Stacking rotations at all stations $\mathcal S=\{r_i\}_{i=1}^R$ and load positions $Z=\{z_k\}_{k=1}^K$ yields the linear system
\begin{equation}
\mathbf y = A\,\mathbf v + \boldsymbol\varepsilon,
\label{eq:yAv}
\end{equation}

with $\mathbf y\in\mathbb R^{RK}$ (measured rotations), $\mathbf v\in\mathbb R^{N}$ (element-averaged compliance), and $A\in\mathbb R^{RK\times N}$ whose entries are
\begin{equation}
\label{eq:A_entries_element}
A_{(r,k),j} := \int_{s_{j-1}}^{s_j} K(r,z_k;s)\,ds 
\;=\; \int_{s_{j-1}}^{s_j} m_r(s)\,M(s;z_k)\,ds.
\end{equation}
Equivalently, one may write $A_{(r,k),j}=\Delta s_j\,\overline{K}_{(r,k),j}$ with the element-average kernel
\begin{equation}
\overline{K}_{(r,k),j} := \frac{1}{\Delta s_j}\int_{s_{j-1}}^{s_j} K(r,z_k;s)\,ds.
\label{eq:Kbar}
\end{equation}
Because $m_r(\cdot)$ and $M(\cdot;z_k)$ are piecewise affine with breakpoints at $s=r$ and $s=z_k$, the integrand in \eqref{eq:A_entries_element} is piecewise quadratic. Elements intersecting these points are split, and the integrals are evaluated analytically. The ultimate goal is to recover $EI(x)=1/v(x)$, or in discrete form $\mathbf{EI}=(1/v_1,\dots,1/v_N)^\top$.

\section{Bayesian Inference Framework}
\label{sec:bayes}

We formulate the discretized forward problem into a statistical model in order to handle measurement noise, prior knowledge, and ill-posedness in a unified manner. Let $\mathbf m\in\mathbb{R}^{N}$ denote the vector of unknown parameters (for example, elementwise compliance $\mathbf v$, flexural stiffness, spring constants, or a reparameterization thereof), and let $\mathbf y\in\mathbb{R}^{M}$ collect the measured responses from all sensor/load combinations, with $M=RK$.

\subsection{Likelihood}
We model the data as the forward response corrupted by additive Gaussian noise with a known correlation structure and unknown scale. Specifically,
\begin{equation}
\label{eq:lik}
\mathbf y \mid \mathbf m,\sigma^{2} \ \sim\ \mathcal N\!\big(F(\mathbf m),\ \sigma^{2}\Gamma\big), 
\qquad
\|u\|_{\Gamma^{-1}}^{2}:=u^\top \Gamma^{-1}u,
\end{equation}
where $u:=F(\mathbf m)-\mathbf y$ denotes the residual vector. Here $F(\mathbf m)$ is the discrete forward map. In the linear case $F(\mathbf m)=A\mathbf m$ with design matrix $A$, so \eqref{eq:lik} reduces to the familiar weighted least–squares setting \autocite{tarantolaInverseProblemTheory2005}. The matrix $\Gamma\succ0$ encodes relative correlations between measurements (for instance across load positions at a fixed sensor), while the scalar $\sigma^2$ scales that correlation to the overall noise level.

\subsection{Priors}
Ill-posedness means many parameter fields can explain the data equally well within noise. A prior distribution regularizes the problem by encoding plausible structure. We primarily use Gaussian priors because they lead to closed-form posteriors in the linear case and to efficient quadratic approximations in the nonlinear case.

A convenient smoothness prior is a \emph{Gaussian Markov random field (GMRF)} centered at $\mathbf{m}_0$ with precision (inverse covariance) $\tau D^\top D$,
\begin{equation}
\label{eq:gmrf}
\mathbf m \mid \tau \;\sim\; 
\mathcal N\!\Big(\mathbf{m}_0,\ \big(\tau\,D^\top D\big)^{-1}\Big), 
\qquad \tau>0.
\end{equation}
Here $D$ is a discrete difference operator (e.g.\ first or second differences, $D_1$ or $D_2$). The operator $D^\top D$ penalizes roughness: larger $\tau$ favors smoother parameter fields that remain close to $\mathbf{m}_0$.

Many structural parameters (e.g.\ flexural rigidity or spring stiffness) are strictly positive. We enforce positivity by introducing a latent variable $\boldsymbol{\eta}$ and writing $\mathbf m=\exp(\boldsymbol{\eta})$ elementwise. The prior is then specified on $\boldsymbol{\eta}$, which is unconstrained on $\mathbb{R}^N$.

Since physical structural properties are expected to vary continuously, we place the GMRF prior directly on the latent field $\boldsymbol{\eta}$:
\begin{equation}
\boldsymbol{\eta} \sim \mathcal N\!\big(\mathbf{m}_{\eta},\ (\tau_{\eta} D^\top D)^{-1}\big).
\end{equation}
This formulation uses the same difference operator $D$ as in \eqref{eq:gmrf} to penalize roughness in $\boldsymbol{\eta}$, and, via the exponential map, produces a strictly positive parameter field $\mathbf m=\exp(\boldsymbol{\eta})$ with spatially correlated variations. The hyperparameter $\tau_{\eta}$ controls the strength of regularization. Because inference is carried out in $\boldsymbol{\eta}$–space, no Jacobian term is needed when optimizing the posterior; the log-normal Jacobian only appears if densities are written in $\mathbf m$–space.

\subsection{Posterior distribution and the MAP–Tikhonov identity}
Combining the likelihood \eqref{eq:lik} with the GMRF prior \eqref{eq:gmrf} yields a Gaussian posterior for the linear model $F(\mathbf m)=A\mathbf m$ \autocite{stuartInverseProblemsBayesian2010,kaipioStatisticalComputationalInverse2006,tarantolaInverseProblemTheory2005,calvettiIntroductionBayesianScientific2007}. Writing:
\begin{equation}
\label{eq:qpost}
Q_{\!\mathrm{post}} = \frac{1}{\sigma^{2}}\,A^\top \Gamma^{-1} A + \tau D^\top D,
\qquad
\mathbf{m}_{\mathrm{post}} = Q_{\!\mathrm{post}}^{-1}
\!\left(\frac{1}{\sigma^{2}}A^\top \Gamma^{-1} \mathbf y + \tau D^\top D\,\mathbf{m}_0\right),
\end{equation}
we have $\mathbf m \mid \mathbf y,\sigma^{2},\tau \sim \mathcal{N}(\mathbf{m}_{\mathrm{post}}, Q_{\!\mathrm{post}}^{-1})$. The negative log-posterior equals a quadratic data misfit plus a quadratic smoothness penalty,
\[
\Phi(\mathbf m;\sigma^{2},\tau)
=\frac{1}{2\sigma^{2}}\|A\mathbf m-\mathbf y\|^{2}_{\Gamma^{-1}}
+\frac{\tau}{2}\|D(\mathbf m-\mathbf{m}_0)\|_{2}^{2},
\]
so the \emph{maximum a posteriori} (MAP) estimate solves a Tikhonov-regularized least-squares problem. The first-order optimality condition,
\[
\big(A^\top \Gamma^{-1} A + \sigma^2\tau D^\top D\big)\,\mathbf m
= A^\top \Gamma^{-1} \mathbf y + \sigma^2\tau D^\top D\,\mathbf{m}_0,
\]
is exactly the normal equation of deterministic \emph{Tikhonov regularization} \autocite{tikhonovSolutionsIllPosed1977,englRegularizationInverseProblems1996,hansenDiscreteInverseProblems2010}. Thus, in the linear-Gaussian setting, the MAP estimator corresponds to the Tikhonov solution with regularization parameter $\lambda=\sigma^{2}\tau$, while the posterior covariance $Q_{\!\mathrm{post}}^{-1}$ supplies uncertainty quantification around that solution.

\subsection{Computation for nonlinear forward models}
When $F$ is nonlinear, we adopt the usual quadratic approximation. At a current iterate $\mathbf{m}_k$ we linearize $F$ as
\[
F(\mathbf m)\approx F(\mathbf{m}_k)+J_k(\mathbf m-\mathbf{m}_k),
\qquad 
J_k:=\frac{\partial F}{\partial \mathbf m}(\mathbf{m}_k),
\qquad
\tilde{\mathbf{y}}_k:=\mathbf y-F(\mathbf{m}_k)+J_k \mathbf{m}_k ,
\]
and replace $A$ by $J_k$ in \eqref{eq:qpost} to obtain a local Gaussian posterior. Iterating these updates yields a \textit{Gauss--Newton} (GN) or \textit{Levenberg--Marquardt} (LM) sequence that converges to the MAP estimator under standard conditions. At the MAP, a Laplace approximation supplies a covariance
\[
\Sigma_{\mathrm{post}}
\;\approx\;
\Big(\frac{1}{\sigma^{2}}J^\top \Gamma^{-1}J + \tau D^\top D\Big)^{-1}\bigg|_{\mathbf{m}=\mathbf{m}_{\mathrm{MAP}}},
\]
which we use to form credible intervals and to propagate uncertainty to derived quantities.

\subsection{Prediction and hyperparameters}
Beyond estimating $\mathbf m$, it is often necessary to predict new measurements under a different design matrix $A_{\mathrm{new}}$ but the same noise model. In the linear case, the \emph{posterior predictive distribution} is Gaussian with mean $A_{\mathrm{new}}\,\mathbf{m}_{\mathrm{post}}$ and covariance $\sigma^{2}\Gamma + A_{\mathrm{new}}\,Q_{\!\mathrm{post}}^{-1}A_{\mathrm{new}}^\top$. For functions of the parameters (for example, flexural rigidity $EI=1/v$ when $\mathbf m=\mathbf v$), we report uncertainty either via the delta method evaluated at $\mathbf{m}_{\mathrm{post}}$ or by sampling from $\mathcal N(\mathbf{m}_{\mathrm{post}},Q_{\!\mathrm{post}}^{-1})$ and mapping through the function.

The noise variance $\sigma^{2}$ and prior precision $\tau$ can be inferred from the data. In an \emph{empirical Bayes} approach, $(\hat\sigma^2,\hat\tau)$ are selected by maximizing the \emph{marginal likelihood} (evidence), which balances data fit and model complexity. In a fully Bayesian treatment, conjugate hyperpriors are assigned, yielding inverse-Gamma posteriors for $\sigma^2$ given $(\mathbf m,\mathbf y)$ and Gamma posteriors for $\tau$ given $\mathbf m$. In both cases, the resulting posterior (or its Laplace approximation in the nonlinear setting) provides parameter estimates together with principled uncertainty quantification.

\section{Sensor informativeness via Fisher information}
\label{subsec:fisher}

Quantifying the \emph{information content} of a sensor at location $r$ regarding the flexural rigidity $EI$ at position $x$ is crucial for optimizing experimental design. Such an assessment guides sensor placement by indicating where measurements are most informative about the flexural rigidity field. We adopt the \emph{Fisher information matrix (FIM)} \autocite{fisherMathematicalFoundationsTheoretical1922,haraldcramerMathematicalMethodsStatistics1946} as a local, model-based measure of identifiability: it quantifies how strongly the data constrain different directions of the parameter vector. Under the Gaussian likelihood in \eqref{eq:lik}, the (expected) FIM is
\begin{equation}
\label{eq:fisher-general}
\mathcal I(\mathbf m)
:= \mathbb E\!\big[\nabla_{\!\mathbf m}\log p(\mathbf y\mid \mathbf m)\,
  \nabla_{\!\mathbf m}\log p(\mathbf y\mid \mathbf m)^\top\big]
= \frac{1}{\sigma^{2}}\,J(\mathbf m)^\top \Gamma^{-1} J(\mathbf m),
\end{equation}
where $J(\mathbf m)=\frac{\partial F}{\partial \mathbf m}(\mathbf{m}) \in \mathbb R^{M\times N}$ denotes the sensitivity (Jacobian) of the forward map. In the linear case, $F(\mathbf m)=A\mathbf m$ so $J(\mathbf m)\equiv A$ and $\mathcal I=\sigma^{-2}A^\top \Gamma^{-1}A$.
Here $M=RK$ is the total number of measurements, with $R$ sensors and $K$ load positions per sensor.

The FIM isolates the \emph{data} contribution to precision, while the \textit{Gaussian Markov random field prior} in \eqref{eq:gmrf} contributes $\tau D^\top D$. In the linear–Gaussian setting this yields the posterior precision as
\begin{equation}
\label{eq:qpost-as-fisher}
Q_{\!\mathrm{post}}
=\underbrace{\tau D^\top D}_{\text{prior precision}}
+\underbrace{\mathcal I}_{\text{data precision}}
\end{equation}
which is exactly \eqref{eq:qpost} rewritten to make the roles of prior and data explicit. In particular, $Q_{\!\mathrm{post}}^{-1}$ reflects the \emph{combined} (data and prior) uncertainty, whereas $\mathcal I$ is prior-agnostic.

Because the measurements are stacked by sensor and load, it is convenient to partition the Jacobian as $J=\big[J_1^\top\,\cdots\,J_R^\top\big]^\top$ with $J_i\in\mathbb R^{K\times N}$ collecting the rows associated with sensor $i$. With the same ordering, the noise covariance is block diagonal, $\Gamma=\mathrm{diag}(\Gamma_1,\dots,\Gamma_R)$, where $\Gamma_i\in\mathbb R^{K\times K}$ stores the correlations for sensor $i$. The information then decomposes additively as:
\begin{equation}
\label{eq:fisher-sum}
\mathcal I
=\frac{1}{\sigma^{2}}\sum_{i=1}^{R} J_i^\top \Gamma_i^{-1}J_i
=:\sum_{i=1}^{R}\mathcal I^{(i)},
\end{equation}
so that $\mathcal I^{(i)}$ is the \emph{per-sensor information}. 

Since the sensitivities are assembled with respect to compliance $\mathbf{v}$ (where $\mathbf{v}$ is the elementwise inverse of rigidity $\mathbf{EI}$), we apply the chain rule to obtain sensitivities with respect to the flexural rigidity parameters $\mathbf{EI}$:
\[
\frac{\partial \mathbf y}{\partial EI_j}
= -\,\frac{1}{EI_j^{2}}\frac{\partial \mathbf y}{\partial v_j}
\qquad \implies \qquad
J_{\cdot,j}^{EI} = -\,EI_j^{-2}\,J_{\cdot,j}^{v},
\]
where $J_{\cdot,j}$ denotes the $j$-th column of the Jacobian. This converts the FIM to the \emph{flexural rigidity parameterization}:
\begin{equation}
\label{eq:fisher-ei}
\mathcal I_{EI}
=\frac{1}{\sigma^{2}}(J^{EI})^\top \Gamma^{-1} J^{EI}
= W^{-1}\!\left(\frac{1}{\sigma^{2}}(J^{v})^\top \Gamma^{-1} J^{v}\right)W^{-1},
\end{equation}
where $W:=\mathrm{diag}(EI_1^{2},\dots,EI_N^{2})$.

\begin{figure}[H]

\begin{subfigure}{\textwidth}
  \centering
  \includegraphics[width=0.75\textwidth]{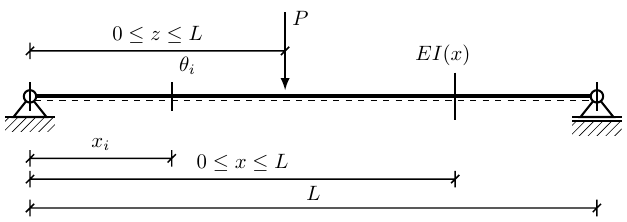}
\end{subfigure}

\begin{subfigure}{\textwidth}
  \centering
  \includegraphics[width=\textwidth]{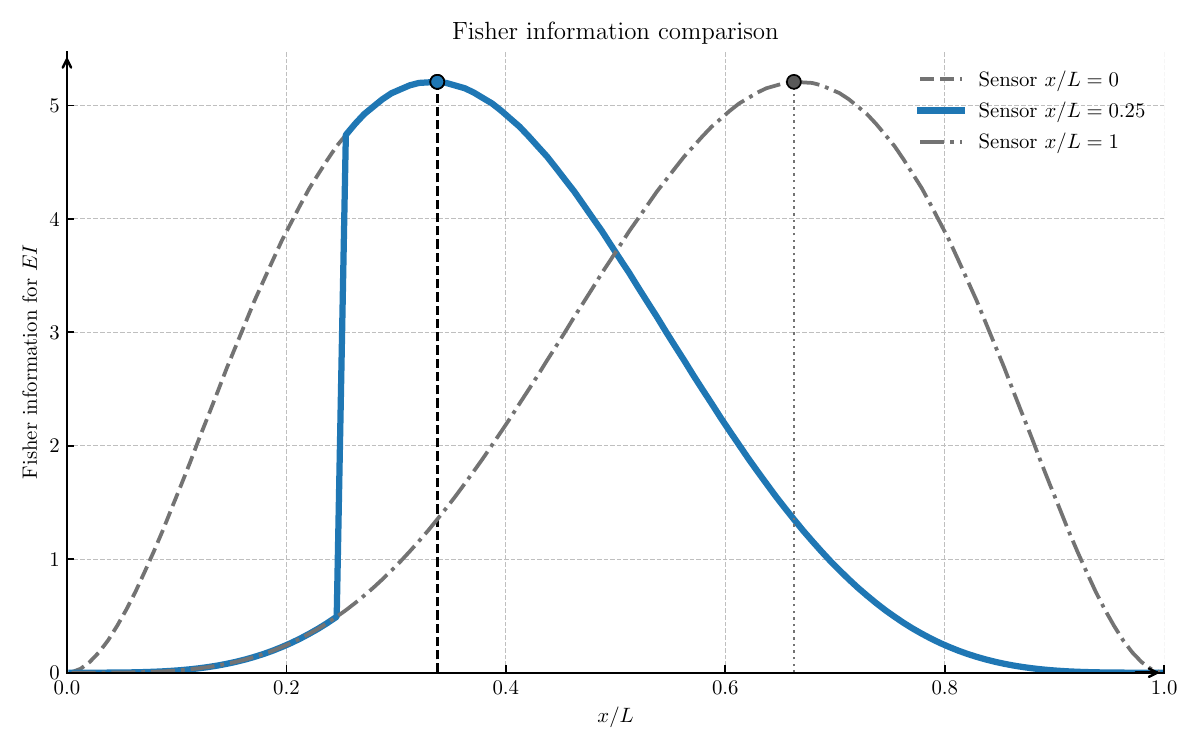}
\end{subfigure}

\caption{Fisher information $\mathcal I^{(i)}_{jj}$ for flexural rigidity $EI$ along the span (computed via \eqref{eq:fi-diag}). Solid blue: sensor at $x/L=1/4$; dashed grey: reference sensors at the supports ($x/L=0$ and $1$). A \emph{kink} occurs at $x/L=1/4$ due to the piecewise kernels $m_r(x)$ and $M(x;\xi)$. Under the adopted load sweep, the blue curve attains its maximum to the right of the sensor; vertical lines mark the \emph{numerically attained} maxima for the plotted curves}
  \label{fig:FisherInformation}
\end{figure}

We illustrate these definitions on a simply supported beam with several alternative sensor locations. The resulting sensor-wise information profiles are reported in \cref{fig:FisherInformation}.

The quantity plotted is the \emph{per–element, per–sensor} diagonal of the Fisher information in the flexural rigidity parameterization. Under independent and identically distributed (i.i.d.) noise ($\Gamma=I$),
\begin{equation}
\label{eq:fi-diag}
\mathcal I^{(i)}_{jj}
=\frac{1}{\sigma^{2}EI_j^{4}}\sum_{k=1}^{K}\big(A_{(i,k),j}\big)^2
\end{equation}
where $A_{(i,k),j}$ is the element integral of the analytic kernel for sensor $r_i$ and load position $z_k$ (see \eqref{eq:A_entries_element}). We refer to the map $x_j\mapsto \mathcal I^{(i)}_{jj}$ as the \emph{informativeness curve} of sensor $i$. Interpreted through the \emph{Cramér–Rao bound} (and ignoring cross-parameter couplings), larger diagonal values correspond to smaller approximate \emph{local} variances for the associated flexural rigidity component.

Two features help interpret these curves. 
First, the scaling is simple: $\mathcal I\propto P^2/\sigma^2$. For fixed span and mesh, each diagonal entry scales with $(\Delta s)^2$ because element sensitivities are integrals over element width.
Second, the \emph{piecewise} nature of the kernel $m_{r_i}(x)M(x;z_k)$ induces visible kinks at sensor locations. With the adopted load sweep, the weighting to the right of a sensor differs from that to the left, so the information a sensor carries about $\mathbf{EI}$ is generally \emph{asymmetric}: For the sensor at \(x/L=1/4\), the informativeness curve remains comparatively low for \(x/L<1/4\), exhibits a discontinuity at the sensor location, attains a peak shortly to the right, and then gradually decays toward the far support. In general, for the simple beam example, the maxima occur near \(x/L=1/3\) for sensors placed at \(x/L<1/3\), and near \(x/L=2/3\) for sensors placed at \(x/L>2/3\).

\begin{figure}[H]
\centering
  \begin{subfigure}{\textwidth}
  \centering
  \includegraphics[width=0.85\textwidth]{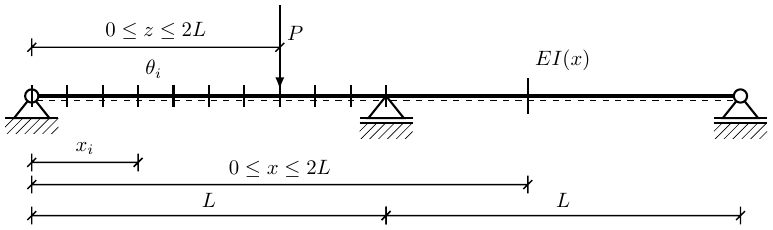}
  \end{subfigure}
  \begin{subfigure}{\textwidth}
  \centering
  \includegraphics[width=\textwidth]{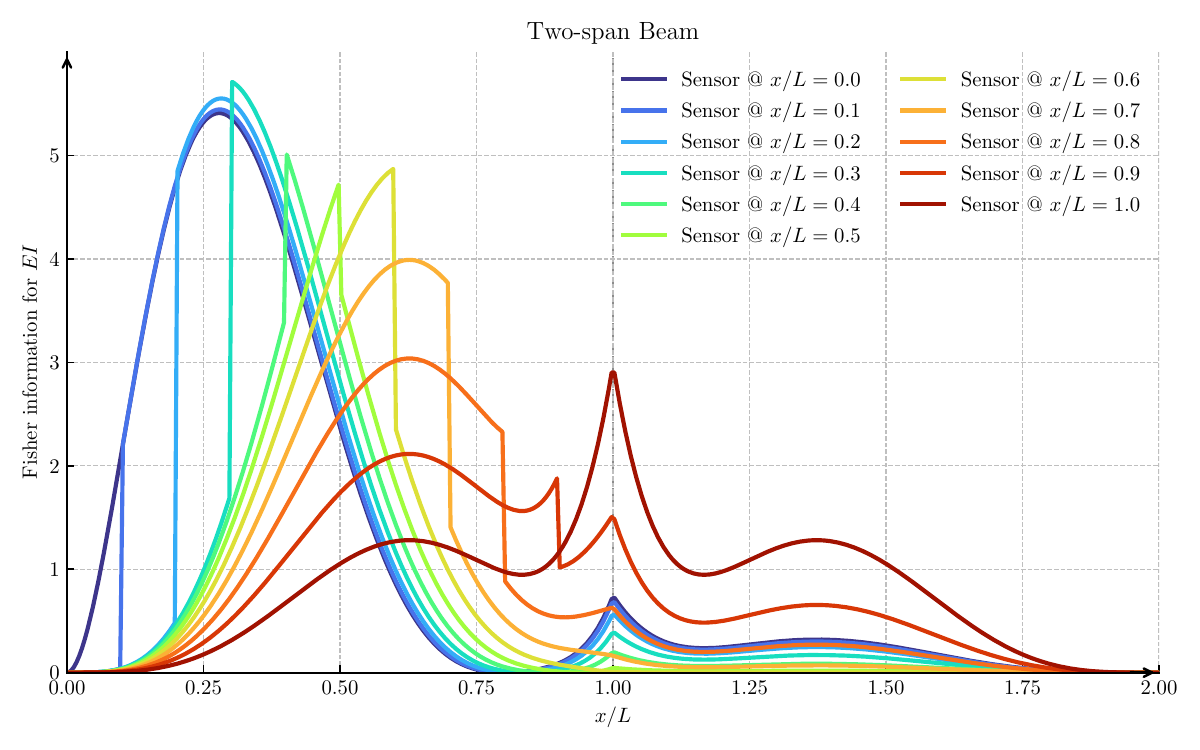}

    \end{subfigure}
  \caption{Fisher information \(\mathcal I^{(i)}_{jj}\) for flexural rigidity \(EI\) along a two-span continuous beam. Each curve corresponds to a single rotation sensor placed on the first span at the indicated position (\(x/L=0,0.1,\dots,1.0\)). The vertical dashed line marks the interior support at \(x/L=1\). Upstream sensors retain diminishing sensitivity beyond the support, while downstream sensors dominate the identifiability of the second span.}
  \label{fig:FisherInformationTwoSpan}

\end{figure}

For a continuous two-span beam (first span \(0\le x/L\le1\), second span \(1\le x/L\le2\)), the same formulation exposes how the interior support redistributes informational sensitivity. \Cref{fig:FisherInformationTwoSpan} reports the Fisher information \(\mathcal I^{(i)}_{jj}\) for flexural rigidity $\mathbf{EI}$ obtained from ten single rotation sensors placed at equally spaced locations along the first span. The first span reproduces the right-skewed asymmetry of the simply supported case, whereas the second span captures the information propagated across the intermediate support. Sensors placed near the midspan of the first span retain measurable, but rapidly diminishing, sensitivity to the adjacent span, while sensors positioned just to the right of the support dominate the local identifiability there. This illustrates how connectivity between spans couples the local sensitivity structure and motivates sensor placement on both sides of interior supports to achieve balanced information coverage.

An eigen-analysis of $\mathcal I$ provides a compact summary of \emph{identifiability}: large eigenvalues correspond to combinations of parameters that are well constrained by the data, whereas small eigenvalues indicate directions that are weakly informed or practically unobservable without additional prior information. Because the linear–Gaussian posterior precision is $Q_{\!\mathrm{post}}=\tau D^\top D+\mathcal I$ (see \eqref{eq:qpost-as-fisher}), the prior acts as a regulariser that supplements missing information and stabilises inversion. In the sense of the \emph{Loewner order}, the posterior covariance always satisfies $Q_{\!\mathrm{post}}^{-1} \preceq \mathcal I^{-1}$ (formally, on the range of $\mathcal I$) since the prior precision is positive semi-definite ($\tau D^\top D \succeq 0$), meaning that adding prior precision can only decrease uncertainty. When $\mathcal I$ is rank-deficient (as occurs with limited sensors), this inequality implies that the prior constrains the unobservable subspace.

For nonlinear forward maps $F$, the expression in \eqref{eq:fisher-general} is evaluated at a nominal parameter (e.g., $\mathbf{m}_0$ or the current Gauss–Newton iterate) using the corresponding Jacobian. The resulting Fisher information is a \emph{local metric} consistent with the quadratic approximation used by the \emph{Gauss–Newton}/\emph{Levenberg–Marquardt} updates and with the \emph{Laplace covariance} evaluated at the maximum a posteriori estimate.

\section{Uncertainty and Identifiability}
\label{sec:uncertainty}

\subsection{Ill-posedness and sources of uncertainty}
The recovery of the flexural \emph{rigidity} from rotation data is an \emph{ill-posed} inverse problem in the sense of \emph{Hadamard}: the forward operator $A$ in \eqref{eq:yAv} is compact and its singular values decay rapidly. Two implications are central for interpretation. First, \emph{noise amplification}: small perturbations in the measured data $\mathbf y$ can induce large changes in the estimated flexural rigidity field. Second, \emph{smoothing (low-pass) behaviour}: fine-scale features of $EI(x)$ are weakly observable and can only be reconstructed with large uncertainty.

Regularisation or Bayesian priors stabilise the inversion but cannot recover information that the data do not carry; they inevitably \emph{trade variance for bias}. In the Bayesian formulation of \S\ref{sec:bayes}, credible bands derived from $Q_{\!\mathrm{post}}$ in \eqref{eq:qpost} quantify the \emph{combined} (data and prior) uncertainty, whereas the Fisher information in \S\ref{subsec:fisher} isolates the \emph{data-driven} component.

Accurate and interpretable flexural rigidity estimation is promoted by a sensing and modelling design that increases information where leverage is strongest. In particular, multiple tilt-sensing stations covering the span mitigate non-identifiability; low sensor noise improves the conditioning of the inverse map; and regularisation aligned with expected behaviour (e.g., second differences for smooth intact beams, edge-preserving when steps/cracks are anticipated) controls variance without introducing undue bias. The Fisher-information analysis in \S\ref{subsec:fisher}, for the simple beam case, further indicates that stations located in the ranges $x/L\in(0,\,1/3)$ and $x/L\in(2/3,\,1)$ maximise information about flexural rigidity near $x/L\approx1/3$ and $x/L\approx2/3$, respectively; using several such stations increases the total information additively, cf.\ \eqref{eq:fisher-sum}.

\subsection{Spatial variation of identifiability}
\begin{figure}[H]
  \centering
  \includegraphics[width=\textwidth]{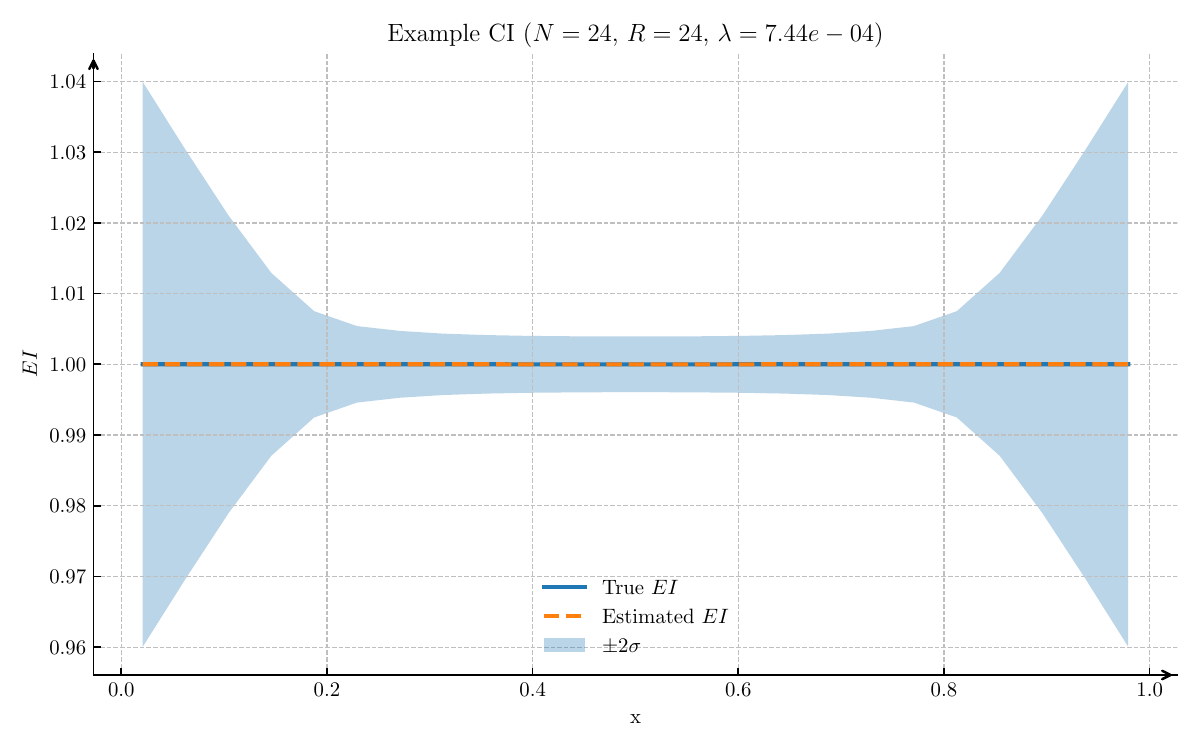}
  \caption{Posterior flexural rigidity profile with \emph{spatially heterogeneous} uncertainty along the span. Solid line: true $EI$; dashed line: posterior mean; blue band: $\pm 2\sigma$ (approx.\ $95\%$ credible interval) from the Bayesian inversion ($N=24$ elements, $K=24$ load positions, $\lambda=7.44\times 10^{-4}$). Uncertainty is largest near the supports and moderately elevated around mid-span, reflecting weaker identifiability where rotations provide less leverage; it narrows in regions where the data are most informative.}
  \label{fig:example_ci}
\end{figure}

On a simply supported span under traversing point loads, the bending moment vanishes at the supports. In the forward relation \eqref{eq:forward}, the integrand $m_r(s)M(s;z)v(s)$ therefore contributes little near the ends across all load positions $z$, so the corresponding columns of $A$ have small norms. As a consequence, the per-element Fisher information \eqref{eq:fi-diag} is minimal near $x=0$ and $x=L$, and the posterior variance is correspondingly large in those regions. Toward mid-span, sensitivities increase and uncertainty narrows. Between widely separated measurement stations, however, distinct flexural rigidity profiles can produce nearly indistinguishable rotation responses, leading to \emph{practical non-identifiability} of the flexural rigidity distribution in those intervals. These spatial patterns of information content are consistent with the per-sensor \emph{informativeness curves} discussed in \S\ref{subsec:fisher} and are illustrated in \cref{fig:example_ci}.

\begin{figure}[H]
\centering
\begin{subfigure}{0.7\textwidth}
\centering
\includegraphics[width=\linewidth]{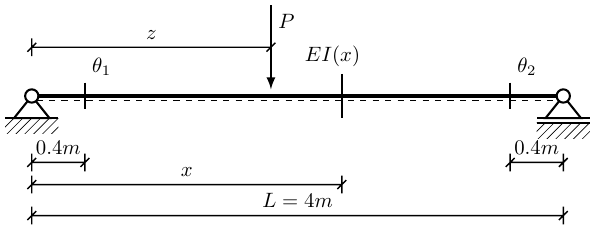}
\end{subfigure}
    \begin{subfigure}{0.9\textwidth}
        \centering
        \includegraphics[width=\linewidth]{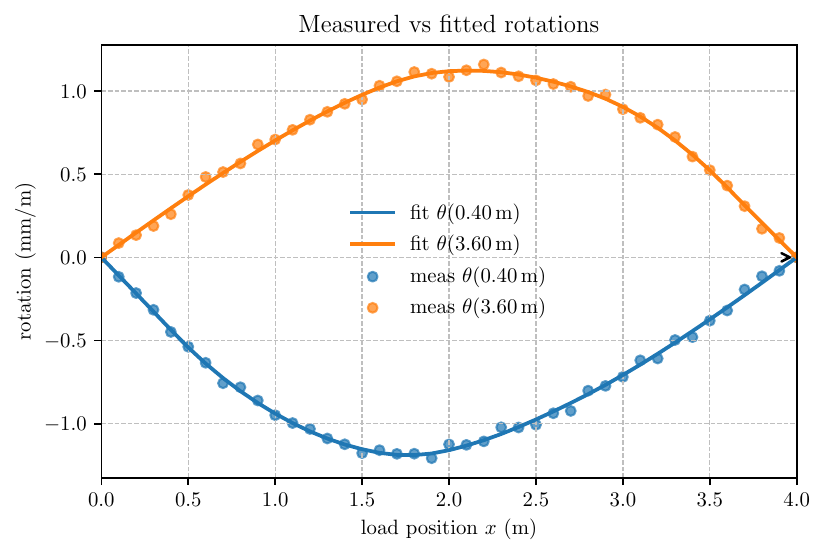}
    \end{subfigure}
            \caption{Rotations at two stations under a traversing point load; tilt-noise s.d. $\sigma=0.02$~mm/m.}
        \label{fig:rot_fit}
\end{figure}

\begin{figure}[H]
    \centering

    \begin{subfigure}{0.48\textwidth}
        \centering
        \includegraphics[width=\linewidth]{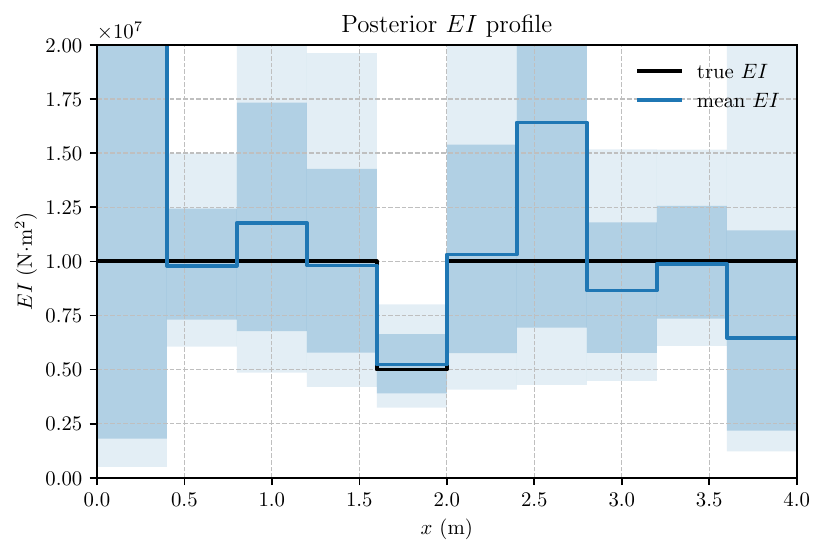}
        \caption{$\sigma=0.02$ mm/m}
        \label{fig:ei_002}
    \end{subfigure}
    \hfill
    \begin{subfigure}{0.48\textwidth}
        \centering
        \includegraphics[width=\linewidth]{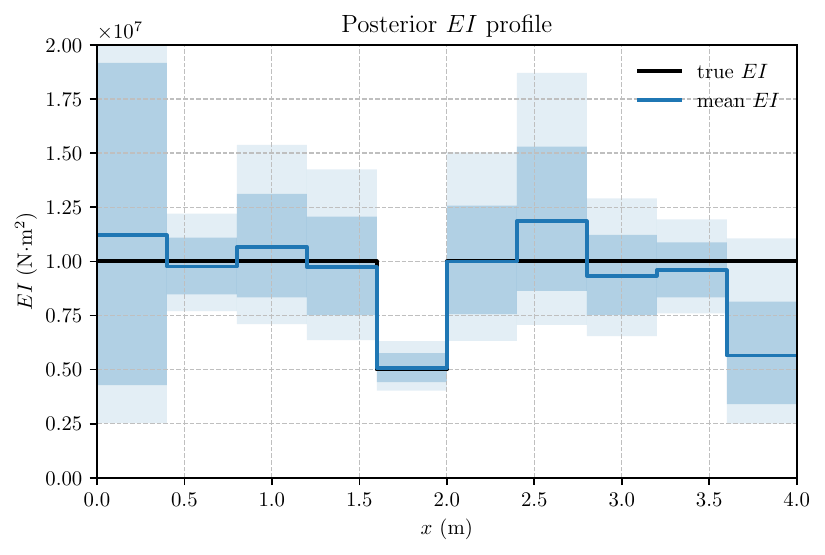}
        \caption{$\sigma=0.01$ mm/m}
        \label{fig:ei_001}
    \end{subfigure}

    \vspace{0.5cm}

    \begin{subfigure}{0.48\textwidth}
        \centering
        \includegraphics[width=\linewidth]{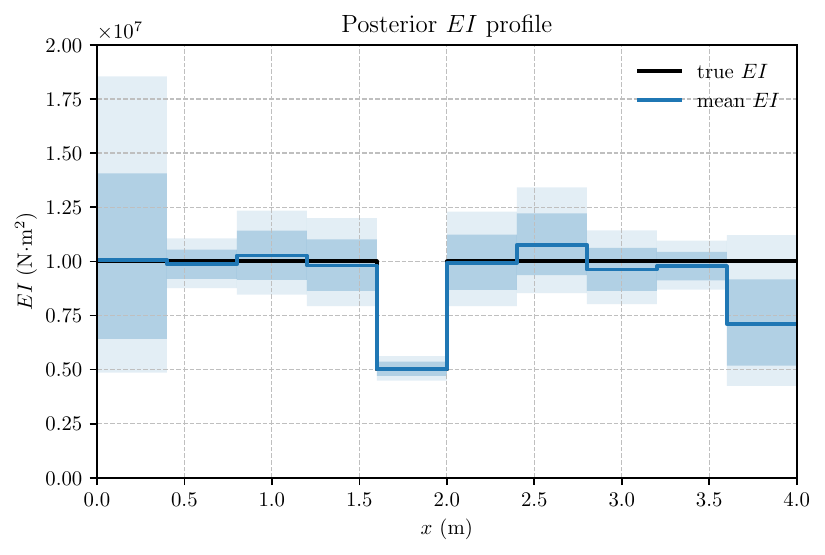}
        \caption{$\sigma=0.005$ mm/m}
        \label{fig:ei_0005}
    \end{subfigure}
    \hfill
    \begin{subfigure}{0.48\textwidth}
        \centering
        \includegraphics[width=\linewidth]{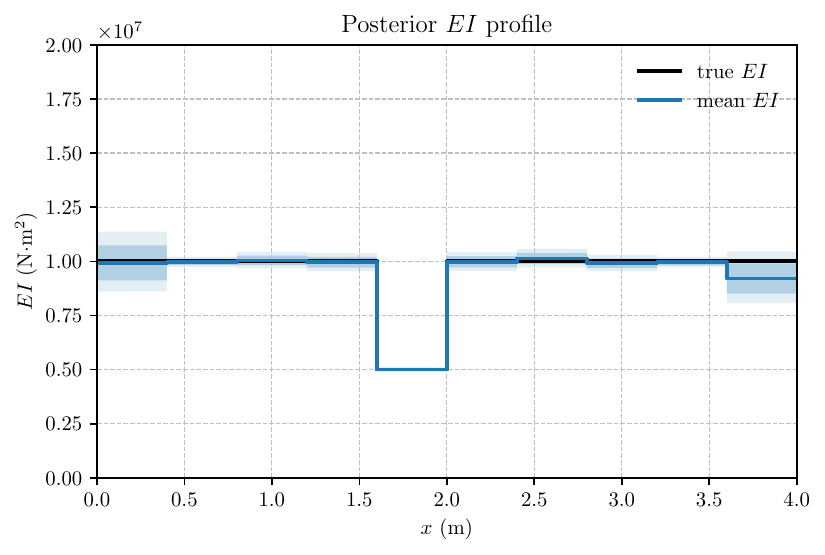}
        \caption{$\sigma=0.001$ mm/m}
        \label{fig:ei_0001}
    \end{subfigure}

    \caption{Bayesian identification of flexural rigidity $EI(x)$ from rotation (tilt) data using 10 elements. Posterior $EI$ profiles for decreasing rotation-noise levels. Black step: ground-truth $EI$ (with an artificial damage zone); blue line: posterior mean; shaded bands: $75\%$ (dark) and $95\%$ (light) credible intervals. As $\sigma$ decreases, uncertainty shrinks and the location/magnitude of the flexural rigidity reduction are recovered more sharply; improvements are most pronounced away from the supports, where the moment field affords greater leverage on $EI$.}
    \label{fig:matrix}
\end{figure}

To further expose the role of measurement quality, \cref{fig:rot_fit,fig:matrix} presents a synthetic study of a locally damaged simply supported beam with two rotation stations and decreasing tilt-noise levels. As the noise standard deviation $\sigma$ is reduced, the posterior bands contract and the damaged zone (reduced flexural rigidity) is recovered more sharply. The improvement is most pronounced away from the supports, where the moment field provides greater leverage on the flexural rigidity; near the ends, uncertainty remains comparatively large even at low noise, in line with the Fisher-information analysis.

\subsection{Bias--Variance Trade-off in Flexural-Rigidity Recovery}
\label{Biasvariance}

For any fixed discretization $N$, number of stations $R$, and regularization weight $\lambda$ (recall $\lambda=\sigma^2\tau$), the total mean–squared error at mid-span ($x=L/2$) decomposes into a \emph{variance} term and a \emph{bias} term. Writing $\widehat{EI}_{1/2}=\widehat{EI}(L/2)$ and letting $EI^{\star}(x)$ denote the true flexural rigidity,
\begin{equation}
\label{eq:bv-rmse}
\mathrm{RMSE}^2\big(\widehat{EI}_{1/2}\big)
\;=\;
\underbrace{\mathrm{Var}\big[\widehat{EI}_{1/2}\big]}_{\textit{variance}}
\;+\;
\underbrace{\Big(\mathbb E[\widehat{EI}_{1/2}] - EI^{\star}(L/2)\Big)^2}_{\textit{bias}^2}
\end{equation}
Expectations and variances are with respect to the data/noise model in \S\ref{sec:bayes}. The variance is inherited from the posterior (or its Laplace/linear analogue in \S\ref{sec:bayes}); the bias collects the systematic effects of \emph{regularization} (shrinkage toward the prior and the null space of $D_2$) and \emph{discretization} (piecewise-constant elements).

\begin{figure}[H]
  \centering
  \includegraphics[width=\textwidth]{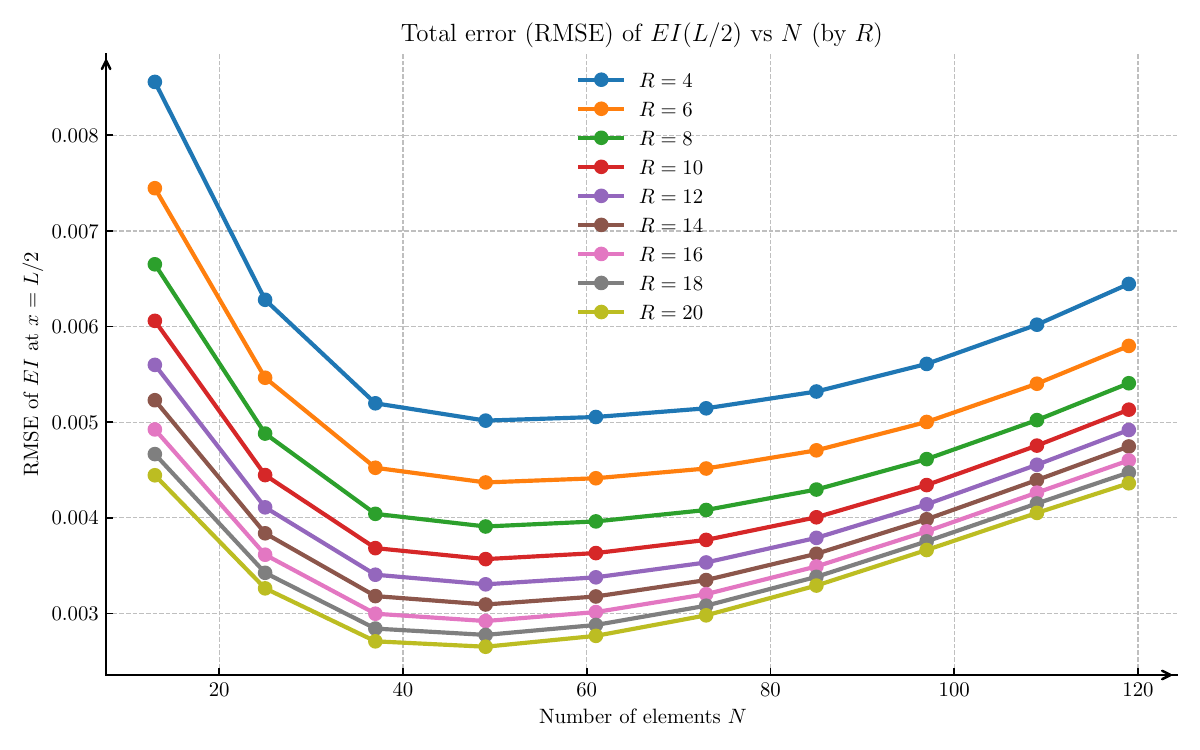}
  \caption{Total error of $EI(L/2)$ versus discretization $N$ for different numbers of rotation sensors $R$. Curves exhibit the characteristic bias–variance trade-off described in \eqref{eq:bv-rmse}.}
  \label{fig:rmse-midspan}
\end{figure}

Figure~\ref{fig:rmse-midspan} shows $\mathrm{RMSE}~\!\big(\widehat{EI}_{1/2}\big)$ versus $N$ for several $R$. Three regimes are visible.

\emph{Coarse meshes (small $N$):} the field is under-resolved and the $D_2$ prior enforces strong smoothing. The \textit{bias} term in \eqref{eq:bv-rmse} dominates, so RMSE is large. Refining the mesh reduces discretization bias and the curves fall.

\emph{Balanced resolution ($N\approx40$–$60$):} the decrease stalls and each curve attains a minimum. Here \textit{bias} has been reduced enough that further refinement yields diminishing returns, while the \textit{variance}—set by the data–to–parameter ratio and the conditioning of $A$—is still moderate. This is the classical \emph{bias–variance sweet spot}.

\emph{Over-parameterized meshes (large $N$):} The number of unknowns grows while the data information saturates ($K$ and $R$ are fixed). The system becomes increasingly ill-conditioned as high-frequency modes become unconstrained by the data. Consequently, noise propagation amplifies the \textit{variance} term in \eqref{eq:bv-rmse}, and the RMSE increases accordingly.

Increasing the number of sensors $R$ raises the data precision $\sum_i J_i^\top \Gamma_i^{-1}J_i$ (see \S\ref{subsec:fisher}), which lowers the \textit{variance} across all $N$ and reduces reliance on strong regularization, thereby also trimming \textit{bias}. In Fig.~\ref{fig:rmse-midspan} the curves shift downward nearly in parallel as $R$ grows, and the minimum RMSE occurs in the same $N$ band: discretization bias is governed by element size, so added sensors chiefly compress the variance without moving the bias-controlled optimum far to the right.

In Fig.~\ref{fig:rmse-midspan}, $\lambda$ is selected once per $R$ by the \emph{Quasi-Optimality} criterion and then reused along the $N$–sweep. This mimics practice (a single tuning per instrumentation layout). If $\lambda$ were re-tuned at every $(N,R)$, the right-hand rise would be partially damped (more regularization at larger $N$), but the qualitative U-shape set by \eqref{eq:bv-rmse} would remain.

The mid-span error is governed by the balance in \eqref{eq:bv-rmse}: refine until discretization \textit{bias} no longer dominates; beyond that point, without additional information (larger $R$ or richer load sets), the \textit{variance} necessarily grows and RMSE worsens. This provides a practical criterion for choosing $N$ given a sensing plan.

\section{Field application: openLAB research bridge}
\label{sec:openlab}

\begin{figure}[H]
    \centering

    \begin{subfigure}{0.69\textwidth}
        \centering
        \includegraphics[width=\linewidth]{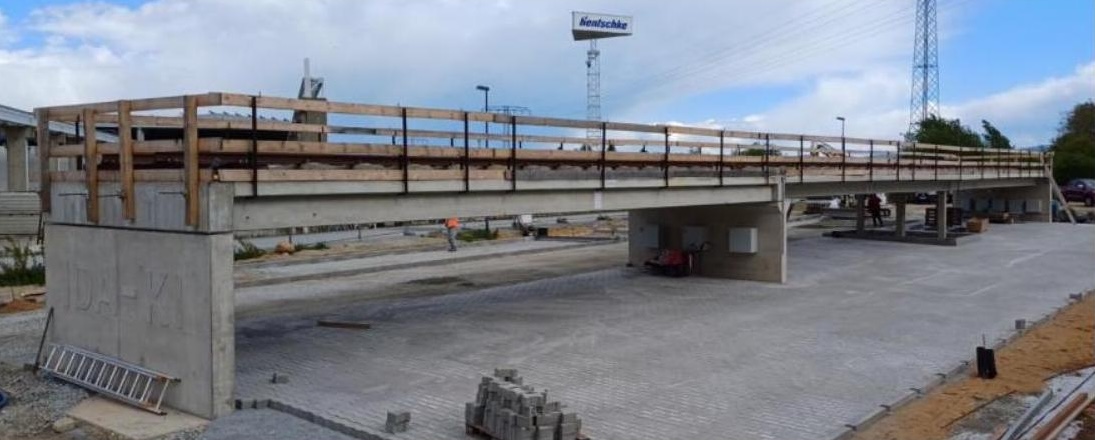}
        \caption{}
    \end{subfigure}
    \hfill
    \begin{subfigure}{0.281\textwidth}
        \centering
        \includegraphics[width=\linewidth]{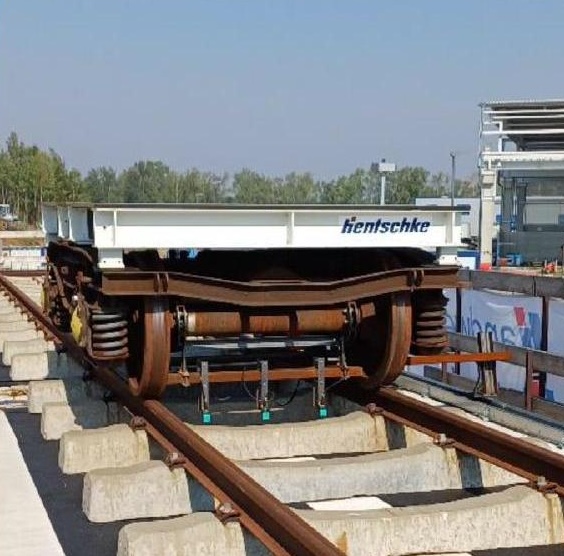}
        \caption{}
    \end{subfigure}
    \caption{The openLAB bridge configurations. (a) The bridge shortly before construction completion in April 2024 \autocite{herbersOpenLABLargescaleDemonstrator2024}; (b) Rail guided load vehicle \autocite{jansenMonitoringDataOpenLAB2025a}; (Photos: Hentschke Bau GmbH).}
    \label{fig:openlab}
\end{figure}

\begin{figure}[H]
  \centering
  \includegraphics[width=\textwidth]{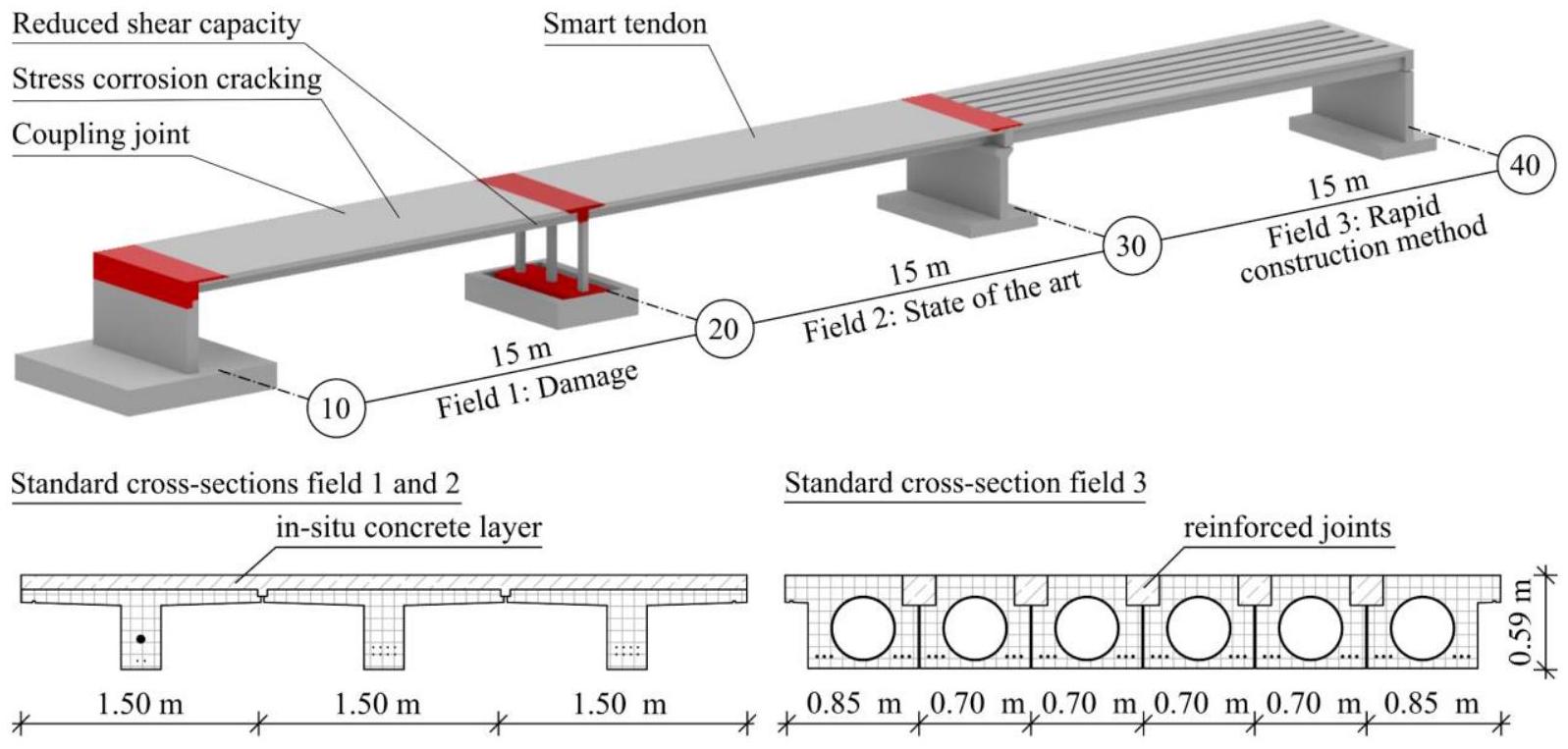}
  \caption{Illustration of the openLAB bridge including the standard cross-sections (Graphic: Fabian Collin, Max Herbers) \autocite{herbersOpenLABLargescaleDemonstrator2024}}
  \label{fig:openlabIllustration}
\end{figure}

The \emph{openLAB bridge} in Bautzen (Germany) is a three-span, 45\,m long and 4.5\,m wide prestressed-concrete demonstrator constructed to support research on SHM and digital twins \autocite{herbersOpenLABLargescaleDemonstrator2024,herbersOpenLABForschungsbrueckeZur2024a,bartelsDocumentationDatasetAcceleration2024,jansenMonitoringDataOpenLAB2025a,herbersMonitoringDataOpenLAB2025}. A rail-guided vehicle with two axles (2\,m spacing) and total mass 4.9\,t provides repeatable crossing loads across the deck. These features are illustrated in \cref{fig:openlab,fig:openlabIllustration}. The present analysis employs data from the curated reference condition release and its supplementary documentation \autocite{jansenMonitoringDataOpenLAB2025a,jansenMonitoringDataOpenLAB2025}. 

\begin{figure}[htbp]
\centering

 \includegraphics[width=0.95\textwidth]{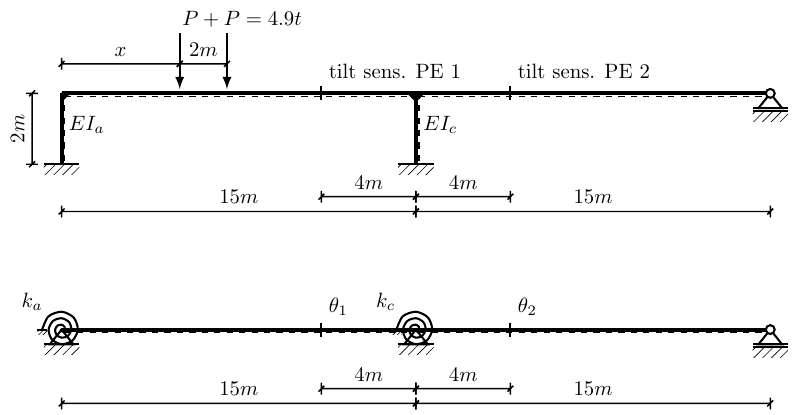}
  \caption{Effective two-span frame model used for inversion (spans~1–2), showing tiltmeter stations, traversing load path, and boundary conditions. The abstraction reflects the construction-induced system transitions described in the data paper (cf.\ Fig.~7 in \autocite{jansenMonitoringDataOpenLAB2025a}).}
\label{fig:FrameSystem}
\end{figure}

For the present study, identification is restricted to the two-span subsystem (spans~1–2). Each of these spans consists of three T-girders with a cast-in-place deck, while span~3 is a slab system. Monolithic connections at axes~10 and~20, in combination with elastomeric bearings at axis~30, result in a frame-like global behaviour for spans~1–2 (\cref{fig:FrameSystem}). Span~3 remains statically independent \autocite{herbersOpenLABLargescaleDemonstrator2024,herbersOpenLABForschungsbrueckeZur2024a}. Six digital tiltmeters are installed, one per girder on spans~1 and~2. Each device is positioned 4\,m from axis~20, close to the zero-moment section under self-weight \autocite{jansenMonitoringDataOpenLAB2025a,jansenMonitoringDataOpenLAB2025}. To simulate traffic actions during the one-year reference phase, the rail-guided vehicle repeatedly crossed the bridge. Tilt data were recorded at 5\,Hz and pre-processed by subtracting the initial 4\,s median and removing mis-triggered samples (cross-correlation threshold~0.85). 
Each span carries three tiltmeters (PE\,ij, with $i=$~span~1–3 and $j=$~girder~1–3, numbered left to right; cf.\ Fig.~7 in \autocite{jansenMonitoringDataOpenLAB2025a}). 
To align the measurements with the two-dimensional structural model, the signals from the three sensors on each span were averaged to yield composite channels (PE\,1 and PE\,2). These correspond to the generalized rotations $\theta_1$ and $\theta_2$ in the frame model (\cref{fig:FrameSystem}).

\begin{figure}[H]
  \centering

  \begin{subfigure}[t]{0.9\linewidth}
    \centering
    \includegraphics[width=\linewidth]{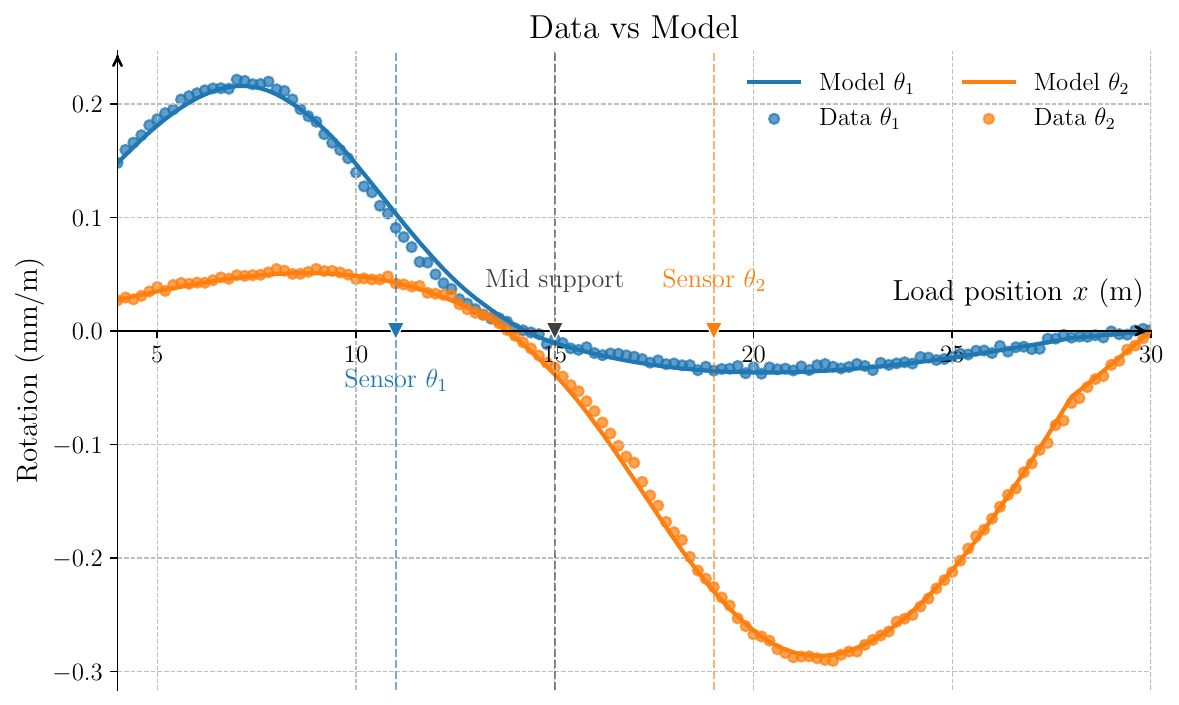}
    \subcaption{6 elements}
    \label{fig:data_vs_model_6E}
  \end{subfigure}

  \medskip

  \begin{subfigure}[t]{0.9\linewidth}
    \centering
    \includegraphics[width=\linewidth]{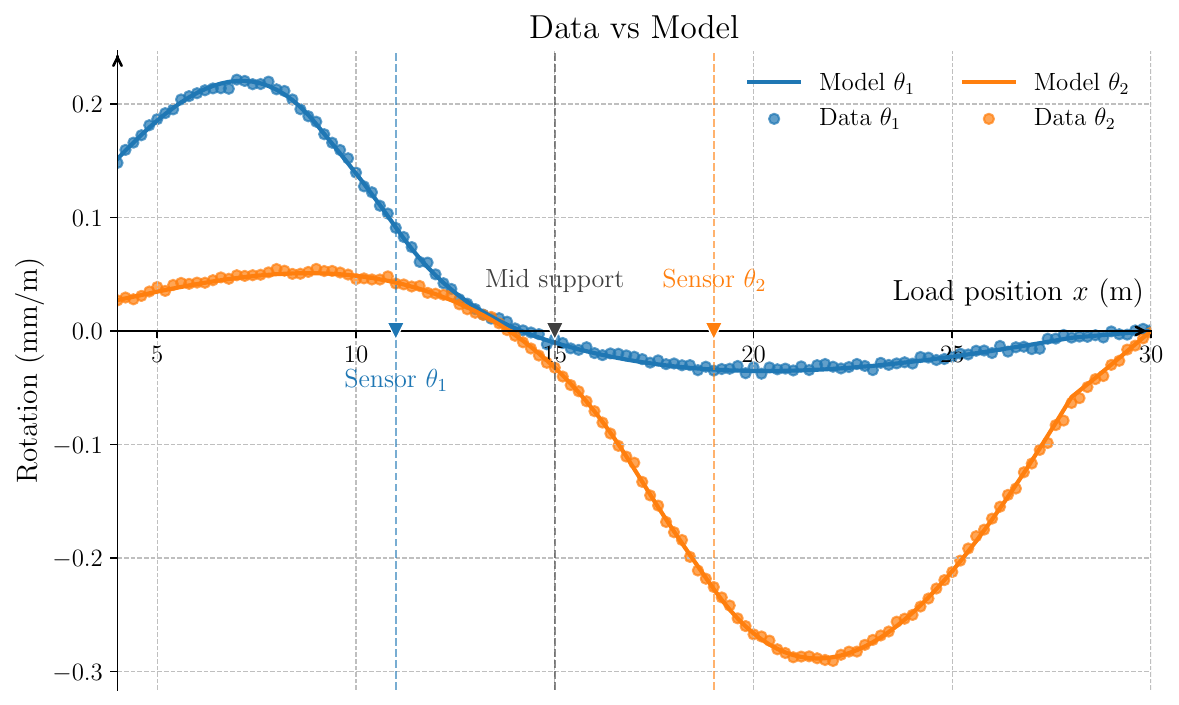}
    \subcaption{30 elements}
    \label{fig:data_vs_model_30E}
  \end{subfigure}

  \caption{Data–model comparison of rotation influence lines at PE1 (blue) and PE2 (orange). Solid curves: model predictions; markers: measurements. $x$ is the load position along the track.}
  \label{fig:data_vs_model}
\end{figure}

Each vehicle crossing thus provides one rotation–time trace per span.
For inversion, only the constant-speed segment (5–30 m along the track) was retained to ensure a direct mapping between time and load position and to avoid acceleration artefacts (\Cref{fig:data_vs_model}).

The forward map follows \S\ref{sec:1}, with measurements stacked over the two stations and discrete load positions. We infer the distributed \emph{flexural rigidity} $EI(x)$ on each span from the \emph{rotation influence lines} using the Bayesian formulation of \S\ref{sec:bayes}. Because the structural system is statically indeterminate, boundary \emph{rotational springs} at the abutments are included and estimated jointly with $EI(x)$ via evidence maximisation (empirical Bayes), thereby avoiding bias in the interior rigidity while acknowledging additional uncertainty from the enlarged parameter set. Uncertainty is reported through posterior credible intervals, obtained from the Gaussian posterior in the linear case or its Laplace approximation in the mildly nonlinear case.

\begin{figure}[H]
  \centering
  \includegraphics[width=\linewidth]{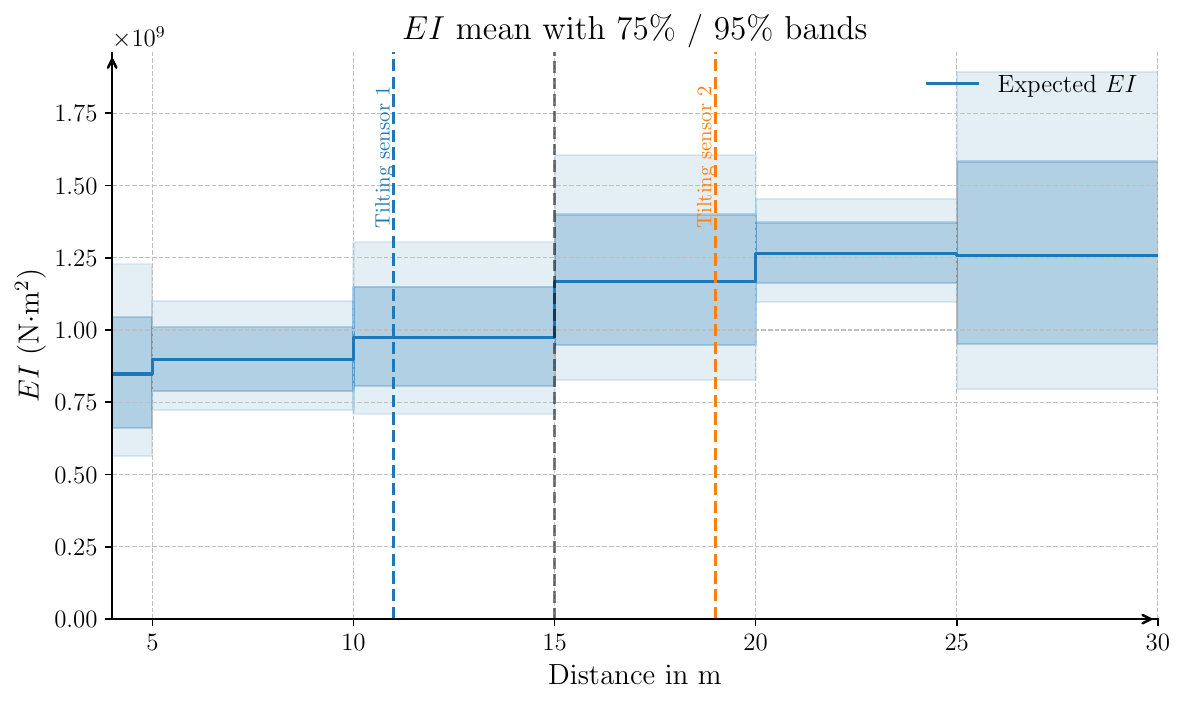}
  \caption{Reconstruction of flexural rigidity $EI(x)$ with 6 elements. Posterior mean with $75\%/95\%$ credible intervals; span~1 is identified as less stiff than span~2, consistent with material differences.}
  \label{fig:EI:distribution:6}
\end{figure}

\Cref{fig:EI:distribution:6} shows the reconstruction using six piecewise–constant elements. Each $EI$ value therefore represents an element average, and $75\%$ and $95\%$ credible bands quantify uncertainty. The posterior mean indicates lower flexural rigidity in span~1 than in span~2, consistent with construction records. While both spans possess identical cross-sections, Span~1 was cast with concrete strength class C25/30 and Span~2 with C50/60, implying a lower elastic modulus for the former. Crucially, the uncertainty is spatially heterogeneous: the bands remain narrow in the span interiors but widen significantly near the zero-moment zones, reflecting reduced sensor informativeness in those regions.

\begin{figure}[H]
  \centering
  \includegraphics[width=\linewidth]{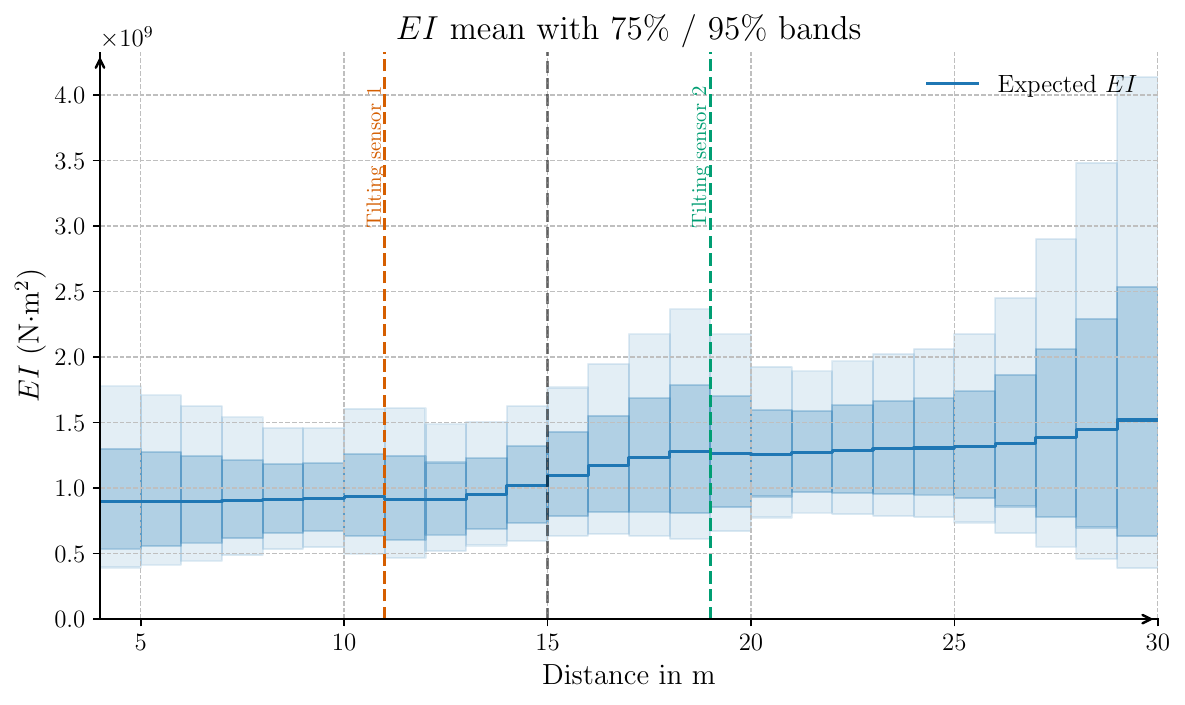}
  \caption{Reconstruction of $EI(x)$ with $\sim$1\,m discretisation. Estimates align with the coarse model, but uncertainty widens, especially near zero-moment region (right-side support), illustrating the identifiability limits discussed in \S\ref{sec:uncertainty}.}
  \label{fig:EI:distribution:32}
\end{figure}

To explore spatial resolution, we refine to approximately one element per metre (\cref{fig:EI:distribution:32}). The posterior mean remains consistent with the six–element result, while credible intervals widen, particularly near $x\!\approx\!30$\,m where bending moments vanish. This pattern is fully aligned with the identifiability analysis in \S\ref{subsec:fisher} and \S\ref{sec:uncertainty}, where the moment field provides little leverage, the corresponding columns of the design matrix are weak and the Fisher information is small, so posterior variance necessarily increases. At fixed instrumentation, finer meshes reduce discretisation bias but increase variance, exemplifying the bias–variance trade–off analysed in \S\ref{Biasvariance}.

The openLAB application highlights three practical points. First, tilt sensing under controlled traversing loads carries sufficient information to recover meaningful spatial variation in flexural rigidity at the span scale, together with principled uncertainty quantification. Second, regions of low bending moment (zero–moment sections) are intrinsically weakly identifiable from rotation data alone, which must be reflected in wider credible intervals. Third, pushing spatial resolution without enriching the data (additional tilt stations and/or complementary modalities such as displacement or dynamic response) leads to variance–dominated reconstructions. In short, the Bayesian framework unifies estimation of $EI(x)$ with uncertainty quantification and provides information–based guidance for instrumentation and load–planning on real bridges.

\section{Conclusion}
This work formulated the identification of distributed flexural rigidity from rotation influence lines as a Bayesian inverse problem. The approach couples an Euler--Bernoulli forward model with elementwise discretisation, a Gaussian likelihood, and smoothness priors that subsume classical \emph{Tikhonov regularization}. Analytical kernel integration and per-sensor Fisher-information diagnostics facilitate efficient computation, while Laplace approximations in nonlinear regimes provide tractable uncertainty quantification, enabling principled credible intervals, sensitivity audits, and bias--variance analyses. Synthetic studies demonstrated how noise level, mesh resolution, and sensor placement jointly govern identifiability. In the openLAB application, the framework recovered span-wise contrasts in flexural rigidity and quantified spatially varying uncertainty under repeatable axle passages.

By embedding regularization in a probabilistic framework, deterministic estimates are elevated to posterior distributions that make explicit how data and prior structure trade off in ill-posed settings. The Fisher-information decomposition isolates the directional content of rotation measurements, revealing asymmetric leverage around zero-moment sections and providing quantitative guidance for instrumentation upgrades. In the openLAB case, posterior bands separate material classes between spans and flag weakly informed regions near supports, supplying transparent evidence for load-rating updates and digital-twin calibration.

The formulation rests on simplifying assumptions: small-deflection Euler--Bernoulli kinematics, quasi-static loading, and log-normal priors that enforce positivity and spatial smoothness. The investigation assimilated data from only two tilt channels and omitted dynamic effects, temperature-induced drifts, torsional coupling, and explicit model-discrepancy terms. Furthermore, the empirical-Bayes tuning relies on the assumption of well-characterized noise correlations. These simplifications constrain resolution in low-moment regions and may underestimate the epistemic uncertainty associated with boundary conditions and modeling idealisations.

Future developments will extend the present framework by (i) fusing rotations with complementary sensing modalities such as displacement or strain influence lines, accelerations, and modal data; (ii) adopting hierarchical priors that infer hyperparameters from ensembles of crossings; (iii) introducing explicit discrepancy models and heavy-tailed noise formulations to enhance robustness; and (iv) implementing online inference for sequential updating as additional passages are recorded. Coupling the Bayesian solver with three-dimensional finite-element surrogates and controlled damage campaigns at openLAB or comparable testbeds will further consolidate validation and support decision-making in maintenance planning, resilience assessment, and value-of-information analyses. Within this trajectory, Bayesian inversion equipped with explicit information metrics offers a scalable route toward uncertainty-aware structural health monitoring and digital-twin development for existing bridge infrastructure.

\section*{Acknowledgement}
The authors gratefully acknowledge the openLAB research team at TU Dresden for providing the experimental data used in this work.
The authors would like to gratefully acknowledge the funding by the Federal Ministry for Digital and Transport (BMDV), Germany, within the mFUND program under grant 19FS2013A.

\printbibliography

\end{document}